\documentclass[american]{elsarticle}
\usepackage[T1]{fontenc}
\usepackage[utf8]{inputenc}
\usepackage{amsmath}
\usepackage{amssymb}
\usepackage{wasysym}
\usepackage{esint}

\makeatletter

\newcommand{\lyxmathsym}[1]{\ifmmode\begingroup\def\b@ld{bold}
  \text{\ifx\math@version\b@ld\bfseries\fi#1}\endgroup\else#1\fi}

\usepackage{amsmath}
\allowdisplaybreaks

\makeatletter
\def\ps@pprintTitle{%
 \let\@oddhead\@empty
 \let\@evenhead\@empty
 \let\@oddfoot\@empty
 \let\@evenfoot\@oddfoot}
\makeatother

\makeatother

\usepackage{babel}
\begin{document}
\begin{frontmatter}
\title{Resolving problems with the continuum limit in coherent-state path
integrals}
\author[rvt]{O. Urbański}
\ead{oliwier1459@gmail.com}
\address[rvt]{Faculty of Physics and Astronomy, Adam Mickiewicz University of Poznań,
Uniwersytetu Poznańskiego 2, 61-614 Poznań, Poland}
\begin{abstract}
The paper solves the problem of continuum limit in bosonic thermal
coherent-state path integrals. For this purpose, exact discrete versions
of the path integral are constructed for three different orderings
of the Hamiltonian: normal, anti-normal and symmetric (Weyl order).
Subsequently, their different continuum versions are checked on the
harmonic oscillator, to choose the symmetric ordering as a possibly
correct choice for all polynomial Hamiltonians. Spotted mathematical
subtleties in the simple case serve as a clue to the general solution.
Finally, a general justification for the symmetric order is provided
by deriving the continuum path integral starting from the exact discrete
case using a renormalization procedure in the imaginary time frequency
domain. While the role of Weyl order has already been found, the paper
provides the missing proof of its suitability for every polynomial
Hamiltonian and simplifies the previously established construction
by referring only to creation and annihilation operators (without
position and momentum operators).
\end{abstract}
\end{frontmatter}
\tableofcontents{}

\section{Introduction}

Path integrals provide an especially interesting and useful way of
handling a wide variety of physical phenomena \citep{kleinert2006path}.
Its particular version, namely the bosonic thermal coherent-state
path integral can be used to treat quantum many-body systems \citep{14168}.
A standard example is given by ultracold atoms in optical lattices
for simulating condensed matter or looking for new states of matter
\citep{polak2007quantum,lewenstein2012ultracold}. In the derivation
of the mentioned tool, a transition from a discrete path formulation
to continuous version is often done without a convincing justification.
As pointed out by Bergeron and Wilson with Galitsky \citep{bergeron1992coherent,wilson2011breakdown},
a very subtle mathematical issues arise (even for a harmonic oscillator),
which if not properly handled lead to incorrect results. Later, a
few authors considered the problem \citep{bruckmann2018rigorous,kordas2014coherent}.
The former provided an interesting procedure for constructing the
path integral, but it does not really correspond to the standard way
of thinking about continuum limit, as in \citep{polak2007quantum}.
The second work does arrive at satisfactory prescription, but provides
no justification beyond checking harmonic oscillator and a single-site
Bose-Hubbard Hamiltonian. Conclusion of Kordas et al. \citep{kordas2014coherent}
is supported by this work. A general proof is provided and also, the
construction is simplified (i. e. formulated without the need to convert
creation and annihilation operators to position and momentum and back).

A different valuable viewpoint on the source of arising subtleties
is provided by Rançon \citep{ranccon2020hubbard}, who refers to the
insight from stochastic calculus. However, it is Klauder and Daubechies
\citep{daubechies1985quantum,klauder2010modern} who provide an outstanding
rigorous description of the continuum coherent-state path integral.
Their construction is based on Wiener measure and is different from
the one developed in this paper. A detailed comparison is reserved
for Sec. \ref{sec:Klauder}.

The paper is organized as follows. Section \ref{sec:Possible-constructions}
provides a careful (almost didactic) development of two well-known
constructions of the coherent-state path integral (based on normal
and anti-normal order of the Hamiltonian). It is finished with a more
obscure procedure using Weyl order (in this form possibly derived
for the first time). Section \ref{sec:Tests-on-the} tests continuum
versions coming from each construction of the path integral on the
harmonic oscillator. The observations made there lead indirectly to
the overall solution to the problem. The main action is condensed
in Sec. \ref{sec:Identifying-and-understanding}, which identifies
the correct continuum construction and derives it carefully from the
exact discrete form. Section \ref{sec:Implications-and-conclusion}
summarizes findings and discusses their meaning.

\section{Possible constructions}\label{sec:Possible-constructions}

\subsection{Procedure using normal order}

This approach is presented in \citep{14168}. We consider many bosonic
modes and associated with them creation operators denoted as $a_{i}^{\dag}$.
Then, a coherent state $\left|\mathbf{z}\right\rangle $ can be defined:
\begin{equation}
\left|\mathbf{z}\right\rangle =\left[\prod_{i}\exp\left(z_{i}a_{i}^{\dagger}\right)\right]\left|0\right\rangle ,\label{|z>}
\end{equation}
where $\mathbf{z}=\left(z_{1},\dots,z_{M}\right)^{T}$ is a (column)
vector composed of complex numbers encoding amplitudes in each mode.
$\left|0\right\rangle $ represents the vacuum state. It is straightforward
to show that coherent states are eigenstates of the annihilation operator:
\begin{align}
a_{i}\left|\mathbf{z}\right\rangle  & =\sum_{\mathbf{n}}\frac{\prod_{j}z_{j}^{n_{j}}}{\sqrt{\prod_{j}n_{j}!}}a_{i}\left|\mathbf{n}\right\rangle =\sum_{\mathbf{n}}\frac{\prod_{j}z_{j}^{n_{j}}}{\sqrt{\prod_{j}n_{j}!}}\sqrt{n_{i}}\left|\mathbf{n}^{\prime\left(i\right)}\right\rangle \nonumber \\
 & =\sum_{\mathbf{n}}\frac{z_{i}z_{i}^{n_{i}-1}\prod_{j\neq i}z_{j}^{n_{j}}}{\sqrt{\left(n_{i}-1\right)!}\sqrt{\prod_{j\neq i}n_{j}!}}\left|\mathbf{n}^{\prime\left(i\right)}\right\rangle \nonumber \\
 & =z_{i}\sum_{\mathbf{n}^{\prime}}\frac{\prod_{j}z_{j}^{n_{j}^{\prime}}}{\sqrt{\prod_{j}n_{j}^{\prime}!}}\left|\mathbf{n}^{\prime\left(i\right)}\right\rangle =z_{i}\left|\mathbf{z}\right\rangle .\label{ai|z>}
\end{align}

The chain of equalities from Eq. (\ref{ai|z>}) uses Fock state representation
of the coherent state, which follows directly from expanding the exponent
in Eq. (\ref{|z>}). $\mathbf{n}=\left(n_{1},\dots,n_{M}\right)^{T}$
is a vector encoding boson numbers in each mode in a Fock state $\left|\mathbf{n}\right\rangle $.
$\left|\mathbf{n}^{\prime\left(i\right)}\right\rangle $ denotes a
state obtained from $\left|\mathbf{n}\right\rangle $ by removing
a single boson from the $i$-th mode. Using a well known consequence
of the Baker--Campbell--Hausdorff formula \citep{Hall:371445},
namely:
\begin{equation}
e^{A}e^{B}=e^{B}e^{A}e^{\left[A,B\right]},\;\text{for }\left[A,\left[A,B\right]\right]=\left[B,\left[A,B\right]\right]=0,
\end{equation}
the following overlap is easily calculated:
\begin{equation}
\left.\left\langle \mathbf{z}_{2}\right|\mathbf{z}_{1}\right\rangle =e^{\mathbf{z}_{2}^{\dagger}\mathbf{z}_{1}}.
\end{equation}

Similarly to Eq. (\ref{ai|z>}), Fock basis can be used to prove the
resolution of identity for coherent states. Both Cartesian and polar
representation of a complex number are used, i. e. $z_{i}=x_{i}+\mathrm{i}y_{i}=r_{i}e^{\mathrm{i}\varphi_{i}}$.
\begin{align}
 & \int\left(\prod_{i}\frac{\mathrm{d}x_{i}\mathrm{d}y_{i}}{\pi}\right)\left|\mathbf{z}\right\rangle \left\langle \mathbf{z}\right|e^{-\mathbf{z}^{\dagger}\mathbf{z}}=\nonumber \\
 & =\int\left(\prod_{i}\frac{r_{i}\mathrm{d}r_{i}\mathrm{d}\varphi_{i}}{\pi}\right)\left(\sum_{\mathbf{n}}\frac{\prod_{i}z_{i}^{n_{i}}}{\sqrt{\prod_{i}n_{i}!}}\left|\mathbf{n}\right\rangle \right)\left(\sum_{\mathbf{m}}\frac{\prod_{i}\bar{z}_{i}^{m_{i}}}{\sqrt{\prod_{i}m_{i}!}}\left\langle \mathbf{m}\right|\right)e^{-\sum_{i}r_{i}^{2}}\nonumber \\
 & =\int\left(\prod_{i}\frac{r_{i}\mathrm{d}r_{i}\mathrm{d}\varphi_{i}}{\pi}\right)\left(\sum_{\mathbf{n}}\frac{\prod_{i}r_{i}^{2n_{i}}}{\prod_{i}n_{i}!}\left|\mathbf{n}\right\rangle \left\langle \mathbf{n}\right|\right)e^{-\sum_{i}r_{i}^{2}}\nonumber \\
 & =\sum_{\mathbf{n}}\left|\mathbf{n}\right\rangle \left\langle \mathbf{n}\right|\frac{\int\left(\prod_{i}\mathrm{d}t_{i}\right)\left(\prod_{i}t_{i}^{n_{i}}e^{-t_{i}}\right)}{\prod_{i}n_{i}!}\nonumber \\
 & =\sum_{\mathbf{n}}\left|\mathbf{n}\right\rangle \left\langle \mathbf{n}\right|\frac{\prod_{i}\int\mathrm{d}t_{i}\,t_{i}^{n_{i}}e^{-t_{i}}}{\prod_{i}n_{i}!}\nonumber \\
 & =\sum_{\mathbf{n}}\left|\mathbf{n}\right\rangle \left\langle \mathbf{n}\right|=I.\label{resolution of identity}
\end{align}
Nondiagonal elements $\left|\mathbf{n}\right\rangle \left\langle \mathbf{m}\right|$
died due to the $\varphi$ integration.

Hamiltonian $H\left(\mathbf{a}^{\dagger},\mathbf{a}\right)$ is an
expression of $a_{i}^{\dagger}$ and $a_{i}$ for many different $i$.
We assume that it is normal-ordered, which means that for every $i$,
$a_{i}^{\dagger}$ always appears to the left of $a_{i}$. Thus:
\begin{align}
 & \left\langle \mathbf{z}_{2}\right|H\left(\mathbf{a}^{\dagger},\mathbf{a}\right)\left|\mathbf{z}_{1}\right\rangle =\left\langle \mathbf{z}_{2}\right|H\left(\mathbf{z}_{2}^{\dagger},\mathbf{z}_{1}\right)\left|\mathbf{z}_{1}\right\rangle =H\left(\mathbf{z}_{2}^{\dagger},\mathbf{z}_{1}\right)\left.\left\langle \mathbf{z}_{2}\right|\mathbf{z}_{1}\right\rangle \nonumber \\
 & =H\left(\mathbf{z}_{2}^{\dagger},\mathbf{z}_{1}\right)e^{\mathbf{z}_{2}^{\dagger}\mathbf{z}_{1}}.
\end{align}
To evaluate the partition function $\mathcal{Z}=\mathrm{Tr}\left[e^{-\beta H}\right]$,
its relation to the imaginary time propagator $U\left(\tau\right)=\exp\left(-H\tau/\hslash\right)$
is employed:
\begin{equation}
\mathcal{Z}=\mathrm{Tr}\left[U\left(\beta\hslash\right)\right].
\end{equation}
Then, multiple resolutions of identity (in terms of coherent states)
are sandwiched between short interval imaginary time propagators:
\begin{equation}
\mathcal{Z}=\mathrm{Tr}\left[\prod_{l=0}^{N-1}U\left(\frac{\beta\hslash}{N}\right)\int\mathrm{d}\mathbf{z}_{l}^{\dag}\mathrm{d}\mathbf{z}_{l}\left|\mathbf{z}_{l}\right\rangle \left\langle \mathbf{z}_{l}\right|e^{-\mathbf{z}_{l}^{\dagger}\mathbf{z}_{l}}\right].\label{Z}
\end{equation}
Integration appearing in Eq. (\ref{Z}) should be understood as follows:
\begin{equation}
\int\mathrm{d}\mathbf{z}^{\dag}\mathrm{d}\mathbf{z}\,\left(\cdots\right)=\int\left(\prod_{i}\frac{\mathrm{d}x_{i}\mathrm{d}y_{i}}{\pi}\right)\,\left(\cdots\right),\label{int}
\end{equation}
where $\mathbf{z}=\left(x_{1}+\mathrm{i}y_{1},\dots,x_{M}+\mathrm{i}y_{M}\right)^{T}$
is decomposed into its real and imaginary parts. Rearranging Eq. (\ref{Z})
we get:
\begin{align}
\mathcal{Z} & =\int\left[\mathcal{D}\mathbf{z}^{\dag}\mathcal{D}\mathbf{z}\right]\prod_{l=0}^{N-1}\left\langle \mathbf{z}_{l-1}\right|U\left(\frac{\beta\hslash}{N}\right)\left|\mathbf{z}_{l}\right\rangle e^{-\mathbf{z}_{l}^{\dagger}\mathbf{z}_{l}},\label{Z-1}
\end{align}
where the integration is assumed over all $\mathbf{z}_{0},\dots,\mathbf{z}_{N-1}$
variables in the sense of Eq. (\ref{int}). Also, periodic boundary
conditions in the imaginary time are naturally assumed, i. e. $\mathbf{z}_{-1}=\mathbf{z}_{N-1}$
and $\mathbf{z}_{0}=\mathbf{z}_{N}$. Bearing in mind that finally
a $N\rightarrow\infty$ limit is meant to be taken, we can write:
\begin{align}
\left\langle \mathbf{z}_{l-1}\right|U\left(\frac{\beta\hslash}{N}\right)\left|\mathbf{z}_{l}\right\rangle  & =\left\langle \mathbf{z}_{l-1}\right|\exp\left[-\frac{\beta}{N}H\left(\mathbf{a}^{\dagger},\mathbf{a}\right)\right]\left|\mathbf{z}_{l}\right\rangle \nonumber \\
 & =\left\langle \mathbf{z}_{l-1}\right|\left[1-\frac{\beta}{N}H\left(\mathbf{a}^{\dagger},\mathbf{a}\right)+\mathcal{O}\left(N^{-2}\right)\right]\left|\mathbf{z}_{l}\right\rangle \nonumber \\
 & =\left[1-\frac{\beta}{N}H\left(\mathbf{z}_{l-1}^{\dag},\mathbf{z}_{l}\right)+\mathcal{O}\left(N^{-2}\right)\right]\left.\left\langle \mathbf{z}_{l-1}\right|\mathbf{z}_{l}\right\rangle \nonumber \\
 & =\exp\left[-\frac{\beta}{N}H\left(\mathbf{z}_{l-1}^{\dagger},\mathbf{z}_{l}\right)+\mathcal{O}\left(N^{-2}\right)\right]\left.\left\langle \mathbf{z}_{l-1}\right|\mathbf{z}_{l}\right\rangle \nonumber \\
 & \cong\exp\left[-\frac{\beta}{N}H\left(\mathbf{z}_{l-1}^{\dagger},\mathbf{z}_{l}\right)\right]e^{\mathbf{z}_{l-1}^{\dagger}\mathbf{z}_{l}}.
\end{align}
Inserting this result into Eq. (\ref{Z-1}) gives
\begin{equation}
\mathcal{Z}=\lim_{N\rightarrow\infty}\int\left[\mathcal{D}\mathbf{z}^{\dag}\mathcal{D}\mathbf{z}\right]\prod_{l=0}^{N-1}\exp\left[-\frac{\beta}{N}H\left(\mathbf{z}_{l-1}^{\dagger},\mathbf{z}_{l}\right)\right]e^{\mathbf{z}_{l-1}^{\dagger}\mathbf{z}_{l}}e^{-\mathbf{z}_{l}^{\dagger}\mathbf{z}_{l}},
\end{equation}
which turns into the following form after collecting the exponentials:
\begin{equation}
\mathcal{Z}=\lim_{N\rightarrow\infty}\int\left[\mathcal{D}\mathbf{z}^{\dag}\mathcal{D}\mathbf{z}\right]\exp\left\{ \sum_{l=0}^{N-1}\left[\left(\mathbf{z}_{l-1}-\mathbf{z}_{l}\right)^{\dag}\mathbf{z}_{l}-\frac{\beta}{N}H\left(\mathbf{z}_{l-1}^{\dagger},\mathbf{z}_{l}\right)\right]\right\} .\label{Z-2}
\end{equation}
Since index $l$ in the sum of Eq. (\ref{Z-2}) can be arbitrarily
shifted, $\mathcal{Z}$ can be rewritten as:
\begin{equation}
\mathcal{Z}=\lim_{N\rightarrow\infty}\int\left[\mathcal{D}\mathbf{z}^{\dag}\mathcal{D}\mathbf{z}\right]\exp\left\{ -\sum_{l=0}^{N-1}\left[\mathbf{z}_{l}^{\dag}\left(\mathbf{z}_{l}-\mathbf{z}_{l+1}\right)+\frac{\beta}{N}H\left(\mathbf{z}_{l}^{\dagger},\mathbf{z}_{l+1}\right)\right]\right\} ,\label{Z normal}
\end{equation}
so that the form of the action is identical to that presented in \citep{bruckmann2018rigorous}
(Eq. (1)).

This normal-ordered discrete path integral is often written in the
continuum version as:
\begin{equation}
\mathcal{Z}=\int\left[\mathcal{D}\mathbf{z}^{\dagger}\mathcal{D}\mathbf{z}\right]\exp\left[\int_{0}^{\beta\hslash}\mathrm{d}\tau\,\mathbf{z}^{\dagger}\left(\tau\right)\frac{\partial}{\partial\tau}\mathbf{z}\left(\tau\right)-\frac{1}{\hslash}\int_{0}^{\beta\hslash}\mathrm{d}\tau\,H\left(\mathbf{z}^{\dagger}\left(\tau\right),\mathbf{z}\left(\tau\right)\right)\right]\label{cont}
\end{equation}
or
\begin{equation}
\mathcal{Z}=\int\left[\mathcal{D}\mathbf{z}^{\dagger}\mathcal{D}\mathbf{z}\right]\exp\left[-\int_{0}^{\beta\hslash}\mathrm{d}\tau\,\mathbf{z}^{\dagger}\left(\tau\right)\frac{\partial}{\partial\tau}\mathbf{z}\left(\tau\right)-\frac{1}{\hslash}\int_{0}^{\beta\hslash}\mathrm{d}\tau\,H\left(\mathbf{z}^{\dagger}\left(\tau\right),\mathbf{z}\left(\tau\right)\right)\right],\label{cont conj}
\end{equation}
which is equivalent, since $\mathcal{Z}$ is real and equals its complex
conjugate. Also, $\hslash$ is completely redundant (it is included
for similarity to the real-time path integral) and can be set to $1$.

It is important to emphasize that Eq. (\ref{cont}) is highly suspicious.
Transition from Eq. (\ref{Z normal}) to Eq. (\ref{cont}) is not
a valid mathematical manipulation. Moreover, ``integration over all
functions'' is ambiguous. This point should become clear, when performing
the naive continuum limit in the next section leads to a different
result.

\subsection{Procedure using anti-normal order}

The following construction is presented in \citep{bruckmann2018rigorous}
and is based on the $P$-representation of the Hamiltonian \citep{scully1997quantum}:
\begin{equation}
\hat{H}=\int\mathrm{d}\mathbf{z}^{\dag}\mathrm{d}\mathbf{z}\,h\left(\mathbf{z}^{\dag},\mathbf{z}\right)\left|\mathbf{z}\right\rangle \left\langle \mathbf{z}\right|e^{-\mathbf{z}^{\dagger}\mathbf{z}}.\label{P-rep}
\end{equation}
It is easily shown, that $h\left(\mathbf{z}^{\dag},\mathbf{z}\right)$
can be obtained from the operator Hamiltonian $\hat{H}$ by putting
it in anti-normal order (please note that only the Hamiltonian gets
now the operator hat symbol, to distinguish it from its representations).
For this purpose, let us focus on a single term:
\begin{equation}
\hat{H}=a_{i}a_{j}\cdots a_{k}^{\dag}a_{l}^{\dag}+\cdots
\end{equation}
and see that it is reconstructed by
\begin{equation}
h\left(\mathbf{z}^{\dag},\mathbf{z}\right)=z_{i}z_{j}\cdots\bar{z}_{k}\bar{z}_{l}+\cdots,
\end{equation}
via Eq. (\ref{P-rep}). Using Eqs. (\ref{ai|z>}) and (\ref{resolution of identity})
leads to
\begin{align}
 & \int\mathrm{d}\mathbf{z}^{\dag}\mathrm{d}\mathbf{z}\,z_{i}z_{j}\cdots\bar{z}_{k}\bar{z}_{l}\left|\mathbf{z}\right\rangle \left\langle \mathbf{z}\right|e^{-\mathbf{z}^{\dagger}\mathbf{z}}=\nonumber \\
 & =\left(a_{i}a_{j}\cdots\right)\left(\int\mathrm{d}\mathbf{z}^{\dag}\mathrm{d}\mathbf{z}\,\left|\mathbf{z}\right\rangle \left\langle \mathbf{z}\right|e^{-\mathbf{z}^{\dagger}\mathbf{z}}\right)\left(\cdots a_{k}^{\dag}a_{l}^{\dag}\right)\nonumber \\
 & =a_{i}a_{j}\cdots a_{k}^{\dag}a_{l}^{\dag},
\end{align}
which finishes the argument.

Let $\Delta=\beta/N$ be the imaginary time-step. Then:
\begin{equation}
I-\Delta\hat{H}=\int\mathrm{d}\mathbf{z}^{\dag}\mathrm{d}\mathbf{z}\left[1-\Delta h\left(\mathbf{z}^{\dag},\mathbf{z}\right)\right]\left|\mathbf{z}\right\rangle \left\langle \mathbf{z}\right|e^{-\mathbf{z}^{\dagger}\mathbf{z}}.
\end{equation}
Following a similar idea to that from the previous section, the partition
function can be expanded as:
\begin{align}
\mathcal{Z} & =\mathrm{Tr}\left[\prod_{l=0}^{N-1}e^{-\Delta\hat{H}}\right]\nonumber \\
 & =\mathrm{Tr}\left[\prod_{l=0}^{N-1}\left[I-\Delta\hat{H}+\mathcal{O}\left(\Delta^{2}\right)\right]\right]\nonumber \\
 & =\mathrm{Tr}\left[\prod_{l=0}^{N-1}\int\mathrm{d}\mathbf{z}_{l}^{\dag}\mathrm{d}\mathbf{z}_{l}\left[1-\Delta h\left(\mathbf{z}_{l}^{\dag},\mathbf{z}_{l}\right)+\mathcal{O}\left(\Delta^{2}\right)\right]\left|\mathbf{z}_{l}\right\rangle \left\langle \mathbf{z}_{l}\right|e^{-\mathbf{z}_{l}^{\dagger}\mathbf{z}_{l}}\right]\nonumber \\
 & =\mathrm{Tr}\left[\int\left[\mathcal{D}\mathbf{z}^{\dag}\mathcal{D}\mathbf{z}\right]\prod_{l=0}^{N-1}\left[1-\Delta h\left(\mathbf{z}_{l}^{\dag},\mathbf{z}_{l}\right)+\mathcal{O}\left(\Delta^{2}\right)\right]\left|\mathbf{z}_{l}\right\rangle \left\langle \mathbf{z}_{l}\right|e^{-\mathbf{z}_{l}^{\dagger}\mathbf{z}_{l}}\right]\nonumber \\
 & =\lim_{N\rightarrow\infty}\int\left[\mathcal{D}\mathbf{z}^{\dag}\mathcal{D}\mathbf{z}\right]\left[\prod_{l=0}^{N-1}e^{-\Delta h\left(\mathbf{z}_{l}^{\dag},\mathbf{z}_{l}\right)}e^{-\mathbf{z}_{l}^{\dagger}\mathbf{z}_{l}}\right]\prod_{l=0}^{N-1}\left.\left\langle \mathbf{z}_{l}\right|\mathbf{z}_{l+1}\right\rangle \nonumber \\
 & =\lim_{N\rightarrow\infty}\int\left[\mathcal{D}\mathbf{z}^{\dag}\mathcal{D}\mathbf{z}\right]\prod_{l=0}^{N-1}e^{\mathbf{z}_{l}^{\dagger}\mathbf{z}_{l+1}-\mathbf{z}_{l}^{\dagger}\mathbf{z}_{l}-\Delta h\left(\mathbf{z}_{l}^{\dag},\mathbf{z}_{l}\right)}\nonumber \\
 & =\lim_{N\rightarrow\infty}\int\left[\mathcal{D}\mathbf{z}^{\dag}\mathcal{D}\mathbf{z}\right]\exp\left\{ -\sum_{l=0}^{N-1}\left[\mathbf{z}_{l}^{\dagger}\left(\mathbf{z}_{l}-\mathbf{z}_{l+1}\right)+\Delta h\left(\mathbf{z}_{l}^{\dag},\mathbf{z}_{l}\right)\right]\right\} .\label{Z an}
\end{align}
Obtained action matches Eq. (4) from \citep{bruckmann2018rigorous}.

The similarity between Eqs. (\ref{Z normal}) and (\ref{Z an}) is
striking. The only differences are subtle, but meaningful. In the
previous representation, Hamiltonian was expressed in normal order,
but it mixed complex amplitudes from adjacent time steps. Now, Hamiltonian
does not mix different time steps, but requires anti-normal form.
Since the mentioned orderings can differ significantly, we find that
using $H\left(\mathbf{z}_{l}^{\dagger},\mathbf{z}_{l+1}\right)$ in
Eq. (\ref{Z normal}) is by no means equivalent to $H\left(\mathbf{z}_{l}^{\dagger},\mathbf{z}_{l}\right)$.
This information gets erased when one passes to the continuum version
(Eq. (\ref{cont})), which suggests that the naive continuum limit
step is incorrect. To put it another way, Eqs. (\ref{Z normal}) and
(\ref{Z an}) have the same continuum versions, but with a different
Hamiltonian form. This point can be also found in \citep{klauder2010modern}.

\subsection{Procedure using Weyl order}

The third, naturally expected construction, uses symmetric order,
known also as Weyl order \citep{cite-key}. It was originally defined
for the position and momentum operators and in such a form presented
in \citep{kordas2014coherent}. An initially similar construction
was given in \citep{dos2005new}, but it lacks the final form found
here and involves real-time. Here, we give an exact imaginary-time
construction starting from a symmetrically ordered Hamiltonian written
in terms of creation and annihilation operators.

Following the original idea of Weyl order, a classical monomial consisting
of canonical position $x$ and momentum $p$ is expressed as an arithmetic
average of all its permutations. For example $x^{2}p=\left(x^{2}p+xpx+px^{2}\right)/3$.
Then, promotion to operators is conducted. Given any classical Hamiltonian
$H\left(p,q\right)$ (a polynomial in $p$ and $q$), one can transform
it into an operator by performing the described procedure on its terms.
A passage in the opposite direction is also possible. One needs to
use the canonical commutation relations to obtain the symmetric order.
Starting from monomials of highest degree produces some extra terms
of lower degrees during reordering. This is not a problem, because
successively monomials of more degrees are correctly arranged until
one gets all of them as required. Finally, demotion of operators to
classical quantities is done.

The key insight coming from \citep{cite-key} is that the relation
just described can be written via Fourier transform:
\begin{equation}
\hat{H}\left(\hat{p},\hat{q}\right)=\int\frac{\mathrm{d}x}{2\pi}\frac{\mathrm{d}k}{2\pi}e^{\mathrm{i}\left(x\hat{p}+k\hat{q}\right)}\int\mathrm{d}p\mathrm{d}q\,e^{-\mathrm{i}\left(xp+kq\right)}H\left(p,q\right).\label{Weyl trick}
\end{equation}
If there are $M>1$ modes, each factor of $2\pi$ has to be raised
to the $M$-th power and multiplication in the exponents has to be
understood as a dot product of vectors. Since powers of $x\hat{p}+k\hat{q}$
(appearing after expanding the exponential) are fully symmetric (with
respect to ordering position and momenta operators), symmetric order
of the Hamiltonian is ensured. The Fourier transforms involved should
be understood as formal ones, because standard convergence usually
is not satisfied.

However, typical Hamiltonians in the field of atomic and optical physics
are expressed in terms of creation and annihilation operators. They
can be easily interchanged for position and momenta operators, but
the resulting formulas look differently and may be unfamiliar. It
is of interest whether Weyl order can be provided without exchanging
the creation and annihilation operators for the canonical ones. The
answer is yes and relies on the following observation. Let
\begin{equation}
S\left(O_{1},\cdots,O_{n}\right)=\frac{1}{n!}\sum_{\sigma}O_{\sigma_{1}}\cdots O_{\sigma_{n}}\label{symmetrizer}
\end{equation}
be the so-called symmetrizer of operator products, where the summation
is carried over all permutations. We can introduce notation
\begin{equation}
S\left[O_{1}\cdots O_{n}+\cdots+O_{1}^{\prime}\cdots O_{n^{\prime}}^{\prime}\right]=S\left(O_{1},\cdots,O_{n}\right)+\cdots+S\left(O_{1}^{\prime},\cdots,O_{n^{\prime}}^{\prime}\right),
\end{equation}
where the left-hand-side is to be understood as a shorthand for the
right-hand-side. $S\left[\cdot\right]$ takes in fact $O_{1}$, $\dots$,
$O_{n}$, $O_{1}^{\prime}$, $\dots$, $O_{n}^{\prime}$ as arguments
(not a single one). $S$ is linear with respect to every argument
in the following sense:
\begin{equation}
S\left[\left(\alpha A_{1}+\beta B_{1}\right)\cdots O_{n}+\cdots\right]=S\left[\alpha A_{1}\cdots O_{n}+\beta B_{1}\cdots O_{n}+\cdots\right],\label{lin}
\end{equation}
here presented only for the first one. Let all operators denoted by
$O$ be either position or momentum operators. We say a given expression
$O_{1}\cdots O_{n}+\cdots+O_{1}^{\prime}\cdots O_{n^{\prime}}^{\prime}$
is Weyl-ordered, if
\begin{equation}
O_{1}\cdots O_{n}+\cdots+O_{1}^{\prime}\cdots O_{n^{\prime}}^{\prime}=S\left[O_{1}\cdots O_{n}+\cdots+O_{1}^{\prime}\cdots O_{n^{\prime}}^{\prime}\right].
\end{equation}
Now, using linear relations between the canonical and creation/annihilation
operators, we can write $O=\alpha a^{\dagger}+\beta a$ for every
operator $O$ (using appropriate coefficients and operators). All
brackets emerging on the left-hand-side can be multiplied out giving
rise to a polynomial written in terms of $a^{\dag},a$. The same can
be done on the right-hand-side, but only thanks to the linearity from
Eq. (\ref{lin}). This way it turns out, that the obtained polynomial
is symmetrically ordered in terms of the creation/annihilation operators.
Summarizing, a direct change from $p,q$ to $a^{\dag},a$ in a symmetrically
ordered expression leaves it symmetrically ordered. Therefore, instead
of translating an atomic/optical Hamiltonian to the language of $p,q$,
then Weyl ordering and finally translating back to the $a^{\dag},a$
language, one can provide symmetrization in the original formulation
of the Hamiltonian. This observation was not pointed out in \citep{kordas2014coherent}.

Writing an analog of Eq. (\ref{Weyl trick}) for the creation and
annihilation operators, for many modes, it becomes:
\begin{equation}
\hat{H}\left(\mathbf{a}^{\dagger},\mathbf{a}\right)=\int\mathrm{d}\tilde{\mathbf{z}}^{\dag}\mathrm{d}\tilde{\mathbf{z}}\,e^{\mathrm{i}\left(\mathbf{a}^{\dagger}\tilde{\mathbf{z}}+\tilde{\mathbf{z}}^{\dag}\mathbf{a}\right)}\int\mathrm{d}\mathbf{z}^{\dag}\mathrm{d}\mathbf{z}\,e^{-\mathrm{i}\left(\mathbf{z}^{\dag}\tilde{\mathbf{z}}+\tilde{\mathbf{z}}^{\dag}\mathbf{z}\right)}\mathcal{H}\left(\mathbf{z}^{\dag},\mathbf{z}\right).\label{Weyl}
\end{equation}
The normalization constant is chosen to correctly transform a constant
Hamiltonian. It turns out to be one, which shows convenience of the
convention given by Eq. (\ref{int}). Next, we follow a similar route
to those from preceding sections to evaluate the partition function:
\begin{align}
 & \mathcal{Z}=\mathrm{Tr}\left[\prod_{l=0}^{N-1}e^{-\Delta\hat{H}}\right]\nonumber \\
 & =\mathrm{Tr}\left[\prod_{l=0}^{N-1}\left[I-\Delta\hat{H}+\mathcal{O}\left(\Delta^{2}\right)\right]\right]\nonumber \\
 & =\mathrm{Tr}\left[\prod_{l=0}^{N-1}\int\mathrm{d}\tilde{\mathbf{z}}_{l}^{\dag}\mathrm{d}\tilde{\mathbf{z}}_{l}\,e^{\mathrm{i}\left(\mathbf{a}^{\dagger}\tilde{\mathbf{z}}_{l}+\tilde{\mathbf{z}}_{l}^{\dag}\mathbf{a}\right)}\int\mathrm{d}\mathbf{z}_{l}^{\dag}\mathrm{d}\mathbf{z}_{l}\,e^{-\mathrm{i}\left(\mathbf{z}_{l}^{\dag}\tilde{\mathbf{z}}_{l}+\tilde{\mathbf{z}}_{l}^{\dag}\mathbf{z}_{l}\right)}\right.\nonumber \\
 & \left.\times\left[1-\Delta\mathcal{H}\left(\mathbf{z}_{l}^{\dag},\mathbf{z}_{l}\right)+\mathcal{O}\left(\Delta^{2}\right)\right]\right]\nonumber \\
 & =\int\left[\mathcal{D}\tilde{\mathbf{z}}^{\dag}\mathcal{D}\tilde{\mathbf{z}}\right]\left[\mathcal{D}\mathbf{z}^{\dag}\mathcal{D}\mathbf{z}\right]\prod_{l=0}^{N-1}\left[e^{-\mathrm{i}\left(\mathbf{z}_{l}^{\dag}\tilde{\mathbf{z}}_{l}+\tilde{\mathbf{z}}_{l}^{\dag}\mathbf{z}_{l}\right)}e^{-\Delta\mathcal{H}\left(\mathbf{z}_{l}^{\dag},\mathbf{z}_{l}\right)+\mathcal{O}\left(\Delta^{2}\right)}\right]\nonumber \\
 & \times\mathrm{Tr}\left[\prod_{l=0}^{N-1}e^{\mathrm{i}\left(\mathbf{a}^{\dagger}\tilde{\mathbf{z}}_{l}+\tilde{\mathbf{z}}_{l}^{\dag}\mathbf{a}\right)}\right]\nonumber \\
 & =\lim_{N\rightarrow\infty}\int\left[\mathcal{D}\tilde{\mathbf{z}}^{\dag}\mathcal{D}\tilde{\mathbf{z}}\right]\left[\mathcal{D}\mathbf{z}^{\dag}\mathcal{D}\mathbf{z}\right]\prod_{l=0}^{N-1}\exp\left[-\mathrm{i}\left(\mathbf{z}_{l}^{\dag}\tilde{\mathbf{z}}_{l}+\tilde{\mathbf{z}}_{l}^{\dag}\mathbf{z}_{l}\right)-\Delta\mathcal{H}\left(\mathbf{z}_{l}^{\dag},\mathbf{z}_{l}\right)\right]\nonumber \\
 & \times\mathrm{Tr}\left[\prod_{l=0}^{N-1}e^{\mathrm{i}\left(\mathbf{a}^{\dagger}\tilde{\mathbf{z}}_{l}+\tilde{\mathbf{z}}_{l}^{\dag}\mathbf{a}\right)}\right].\label{symZ}
\end{align}
Now, however, the trace is a little bit more tricky to calculate.
It can be done by employing another path integral:
\begin{align}
 & \mathrm{Tr}\left[\prod_{l=0}^{N-1}e^{\mathrm{i}\left(\mathbf{a}^{\dagger}\tilde{\mathbf{z}}_{l}+\tilde{\mathbf{z}}_{l}^{\dag}\mathbf{a}\right)}\right]=\nonumber \\
 & =\mathrm{Tr}\left[\prod_{l=0}^{N-1}e^{\mathrm{i}\mathbf{a}^{\dagger}\tilde{\mathbf{z}}_{l}}e^{\mathrm{i}\tilde{\mathbf{z}}_{l}^{\dag}\mathbf{a}}e^{-\frac{1}{2}\left[\mathrm{i}\mathbf{a}^{\dagger}\tilde{\mathbf{z}}_{l},\mathrm{i}\tilde{\mathbf{z}}_{l}^{\dag}\mathbf{a}\right]}\right]\nonumber \\
 & =\left(\prod_{l=0}^{N-1}e^{-\frac{1}{2}\tilde{\mathbf{z}}_{l}^{\dag}\tilde{\mathbf{z}}_{l}}\right)\mathrm{Tr}\left[\prod_{l=0}^{N-1}\int\mathrm{d}\mathbf{z}_{l}^{\dag}\mathrm{d}\mathbf{z}_{l}\left|\mathbf{z}_{l}\right\rangle \left\langle \mathbf{z}_{l}\right|e^{-\mathbf{z}_{l}^{\dagger}\mathbf{z}_{l}}e^{\mathrm{i}\mathbf{a}^{\dagger}\tilde{\mathbf{z}}_{l}}e^{\mathrm{i}\tilde{\mathbf{z}}_{l}^{\dag}\mathbf{a}}\right]\nonumber \\
 & =\left(\prod_{l=0}^{N-1}e^{-\frac{1}{2}\tilde{\mathbf{z}}_{l}^{\dag}\tilde{\mathbf{z}}_{l}}\right)\int\left[\mathcal{D}\mathbf{z}^{\dag}\mathcal{D}\mathbf{z}\right]\mathrm{Tr}\left[\prod_{l=0}^{N-1}\left\langle \mathbf{z}_{l}\right|e^{-\mathbf{z}_{l}^{\dagger}\mathbf{z}_{l}}e^{\mathrm{i}\mathbf{a}^{\dagger}\tilde{\mathbf{z}}_{l}}e^{\mathrm{i}\tilde{\mathbf{z}}_{l}^{\dag}\mathbf{a}}\left|\mathbf{z}_{l+1}\right\rangle \right]\nonumber \\
 & =\left(\prod_{l=0}^{N-1}e^{-\frac{1}{2}\tilde{\mathbf{z}}_{l}^{\dag}\tilde{\mathbf{z}}_{l}}\right)\int\left[\mathcal{D}\mathbf{z}^{\dag}\mathcal{D}\mathbf{z}\right]\prod_{l=0}^{N-1}e^{-\mathbf{z}_{l}^{\dagger}\mathbf{z}_{l}}\left\langle \mathbf{z}_{l}\right|e^{\mathrm{i}\mathbf{a}^{\dagger}\tilde{\mathbf{z}}_{l}}e^{\mathrm{i}\tilde{\mathbf{z}}_{l}^{\dag}\mathbf{a}}\left|\mathbf{z}_{l+1}\right\rangle \nonumber \\
 & =\left(\prod_{l=0}^{N-1}e^{-\frac{1}{2}\tilde{\mathbf{z}}_{l}^{\dag}\tilde{\mathbf{z}}_{l}}\right)\int\left[\mathcal{D}\mathbf{z}^{\dag}\mathcal{D}\mathbf{z}\right]\prod_{l=0}^{N-1}e^{-\mathbf{z}_{l}^{\dagger}\mathbf{z}_{l}}e^{\mathrm{i}\mathbf{z}_{l}^{\dag}\tilde{\mathbf{z}}_{l}}e^{\mathrm{i}\tilde{\mathbf{z}}_{l}^{\dag}\mathbf{z}_{l+1}}e^{\mathbf{z}_{l}^{\dag}\mathbf{z}_{l+1}}\nonumber \\
 & =\left(\prod_{l=0}^{N-1}e^{-\frac{1}{2}\tilde{\mathbf{z}}_{l}^{\dag}\tilde{\mathbf{z}}_{l}}\right)\int\left[\mathcal{D}\mathbf{z}^{\dag}\mathcal{D}\mathbf{z}\right]\exp\left\{ \sum_{l=0}^{N-1}\left[\mathbf{z}_{l}^{\dagger}\left(\mathbf{z}_{l+1}-\mathbf{z}_{l}\right)+\mathrm{i}\left(\mathbf{z}_{l}^{\dag}\tilde{\mathbf{z}}_{l}+\tilde{\mathbf{z}}_{l}^{\dag}\mathbf{z}_{l+1}\right)\right]\right\} \nonumber \\
 & =\left(\prod_{l=0}^{N-1}e^{-\frac{1}{2}\tilde{\mathbf{z}}_{l}^{\dag}\tilde{\mathbf{z}}_{l}}\right)\int\left[\mathcal{D}\check{\mathbf{z}}^{\dag}\mathcal{D}\check{\mathbf{z}}\right]\exp\left\{ \sum_{l=0}^{N-1}\left[\check{\mathbf{z}}_{l}^{\dagger}\left(\check{\mathbf{z}}_{l+1}-\check{\mathbf{z}}_{l}\right)+\mathrm{i}\left(\check{\mathbf{z}}_{l}^{\dag}\tilde{\mathbf{z}}_{l}+\tilde{\mathbf{z}}_{l}^{\dag}\check{\mathbf{z}}_{l+1}\right)\right]\right\} .\label{trace}
\end{align}
Inserting Eq. (\ref{trace}) to Eq. (\ref{symZ}) gives
\begin{align}
 & \mathcal{Z}=\lim_{N\rightarrow\infty}\int\left[\mathcal{D}\mathbf{z}^{\dag}\mathcal{D}\mathbf{z}\right]\left[\mathcal{D}\tilde{\mathbf{z}}^{\dag}\mathcal{D}\tilde{\mathbf{z}}\right]\left[\mathcal{D}\check{\mathbf{z}}^{\dag}\mathcal{D}\check{\mathbf{z}}\right]\nonumber \\
 & \times\exp\left\{ \sum_{l=0}^{N-1}\left[\mathrm{i}\left(\check{\mathbf{z}}_{l}^{\dag}\tilde{\mathbf{z}}_{l}+\tilde{\mathbf{z}}_{l}^{\dag}\check{\mathbf{z}}_{l+1}\right)-\mathrm{i}\left(\mathbf{z}_{l}^{\dag}\tilde{\mathbf{z}}_{l}+\tilde{\mathbf{z}}_{l}^{\dag}\mathbf{z}_{l}\right)\right.\right.\nonumber \\
 & \left.\left.-\frac{1}{2}\tilde{\mathbf{z}}_{l}^{\dag}\tilde{\mathbf{z}}_{l}+\check{\mathbf{z}}_{l}^{\dagger}\left(\check{\mathbf{z}}_{l+1}-\check{\mathbf{z}}_{l}\right)-\Delta\mathcal{H}\left(\mathbf{z}_{l}^{\dag},\mathbf{z}_{l}\right)\right]\right\} \nonumber \\
 & =\lim_{N\rightarrow\infty}\int\left[\mathcal{D}\mathbf{z}^{\dag}\mathcal{D}\mathbf{z}\right]f\left(\mathbf{z}^{\dag},\mathbf{z}\right)\exp\left[-\sum_{l=0}^{N-1}\Delta\mathcal{H}\left(\mathbf{z}_{l}^{\dag},\mathbf{z}_{l}\right)\right],
\end{align}
where
\begin{align}
 & f\left(\mathbf{z}^{\dag},\mathbf{z}\right)=\nonumber \\
 & =\int\left[\mathcal{D}\tilde{\mathbf{z}}^{\dag}\mathcal{D}\tilde{\mathbf{z}}\right]\left[\mathcal{D}\check{\mathbf{z}}^{\dag}\mathcal{D}\check{\mathbf{z}}\right]\exp\left\{ \sum_{l=0}^{N-1}\left[\mathrm{i}\left(\check{\mathbf{z}}_{l}^{\dag}\tilde{\mathbf{z}}_{l}+\tilde{\mathbf{z}}_{l}^{\dag}\check{\mathbf{z}}_{l+1}\right)\right.\right.\nonumber \\
 & \left.\left.-\mathrm{i}\left(\mathbf{z}_{l}^{\dag}\tilde{\mathbf{z}}_{l}+\tilde{\mathbf{z}}_{l}^{\dag}\mathbf{z}_{l}\right)-\frac{1}{2}\tilde{\mathbf{z}}_{l}^{\dag}\tilde{\mathbf{z}}_{l}+\check{\mathbf{z}}_{l}^{\dagger}\left(\check{\mathbf{z}}_{l+1}-\check{\mathbf{z}}_{l}\right)\right]\right\} \nonumber \\
 & =\int\left[\mathcal{D}\tilde{\mathbf{z}}^{\dag}\mathcal{D}\tilde{\mathbf{z}}\right]\left[\mathcal{D}\check{\mathbf{z}}^{\dag}\mathcal{D}\check{\mathbf{z}}\right]\exp\left\{ \sum_{ll^{\prime}}\begin{pmatrix}\tilde{\mathbf{z}}_{l}^{\dag} & \check{\mathbf{z}}_{l}^{\dag}\end{pmatrix}\begin{pmatrix}-\frac{1}{2}\delta_{ll^{\prime}} & \mathrm{i}\delta_{l+1,l^{\prime}}\\
\mathrm{i}\delta_{ll^{\prime}} & \delta_{l+1,l^{\prime}}-\delta_{ll^{\prime}}
\end{pmatrix}\begin{pmatrix}\tilde{\mathbf{z}}_{l^{\prime}}\\
\check{\mathbf{z}}_{l^{\prime}}
\end{pmatrix}\right.\nonumber \\
 & \left.-\mathrm{i}\sum_{l}\begin{pmatrix}\mathbf{z}_{l}^{\dag} & \mathrm{0}\end{pmatrix}\begin{pmatrix}\tilde{\mathbf{z}}_{l}\\
\check{\mathbf{z}}_{l}
\end{pmatrix}-\mathrm{i}\sum_{l}\begin{pmatrix}\tilde{\mathbf{z}}_{l}^{\dag} & \mathrm{\check{\mathbf{z}}}_{l}^{\dag}\end{pmatrix}\begin{pmatrix}\mathbf{z}_{l}\\
0
\end{pmatrix}\right\} \nonumber \\
 & =\int\left[\mathcal{D}\tilde{\mathbf{z}}^{\dag}\mathcal{D}\tilde{\mathbf{z}}\right]\left[\mathcal{D}\check{\mathbf{z}}^{\dag}\mathcal{D}\check{\mathbf{z}}\right]\exp\left\{ \begin{pmatrix}\tilde{\mathbf{z}}^{\dag} & \check{\mathbf{z}}^{\dag}\end{pmatrix}\mathbf{M}\begin{pmatrix}\tilde{\mathbf{z}}\\
\check{\mathbf{z}}
\end{pmatrix}+\mathbf{A}^{\dag}\begin{pmatrix}\tilde{\mathbf{z}}\\
\check{\mathbf{z}}
\end{pmatrix}+\begin{pmatrix}\tilde{\mathbf{z}}^{\dag} & \check{\mathbf{z}}^{\dag}\end{pmatrix}\mathbf{B}\right\} \nonumber \\
 & =\frac{\exp\left(-\mathbf{A}^{\dag}\mathbf{M}^{-1}\mathbf{B}\right)}{\det\left(-\mathbf{M}\right)}.\label{f}
\end{align}

To evaluate easily matrix multiplication and the determinant, it is
convenient to work in the the Matsubara frequency domain. Any discrete
sequence $u_{l}$ is related to its frequency representation $u_{\omega}$
via discrete Fourier transform:
\begin{equation}
u_{l}=\frac{1}{\sqrt{N}}\sum_{\omega}u_{\omega}e^{\mathrm{i}\omega l},\;\omega_{n}=\frac{2\pi}{N}n.\label{fourier}
\end{equation}
Identity matrix remains identity ($\delta_{ll^{\prime}}\rightarrow1$)
and a simple ``shifting'' operator $\delta_{l+1,l^{\prime}}$ becomes
$e^{\mathrm{i}\omega}$. In this representation
\begin{equation}
\mathbf{M}=\begin{pmatrix}-\frac{1}{2} & \mathrm{i}e^{\mathrm{i}\omega}\\
\mathrm{i} & e^{\mathrm{i}\omega}-1
\end{pmatrix}\delta_{\omega\omega^{\prime}}\delta_{mm^{\prime}}
\end{equation}
and
\begin{equation}
\mathbf{A}^{\dag}=-\mathrm{i}\begin{pmatrix}\mathbf{z}_{\omega}^{\dag} & \mathrm{0}\end{pmatrix},\;\mathbf{B}=-\mathrm{i}\begin{pmatrix}\mathbf{z}_{\omega}\\
0
\end{pmatrix},
\end{equation}
where indices $m,m^{\prime}$ refer to different modes. Determinant
of $-\mathbf{M}$ is a product of single $2\times2$ matrix determinants
for each frequency and mode:
\begin{equation}
\det\left(-\mathbf{M}\right)=\left[\prod_{\omega}\frac{1}{2}\left(1+e^{\mathrm{i}\omega}\right)\right]^{M}.
\end{equation}
The product is taken over all distinct (modulo $2\pi$) frequencies
of the form $2\pi n/N$ and can be evaluated by means of the following
trick. Since $e^{\mathrm{i}\omega}$ runs over all $N$-th degree
roots of unity, then factoring a simple polynomial $z^{N}-1$ produces
a similar product:
\begin{equation}
z^{N}-1=\prod_{\omega}\left(z-e^{\mathrm{i}\omega}\right).
\end{equation}
Setting $z=-1$ gives
\begin{equation}
\prod_{\omega}\frac{1}{2}\left(1+e^{\mathrm{i}\omega}\right)=\begin{cases}
2^{1-N} & \text{for odd }N\\
0 & \text{for even }N
\end{cases}.\label{prod identity}
\end{equation}
For even $N$ the determinant (which appears in the denominator of
Eq. (\ref{f})) vanishes and the construction simply fails. However,
it can be successfully completed for odd $N$, which finally does
not matter, because $N\rightarrow\infty$. Inversion of matrix $\mathbf{M}$
leads to
\begin{align}
-\mathbf{A}^{\dag}\mathbf{M}^{-1}\mathbf{B} & =\sum_{\omega}\begin{pmatrix}\mathbf{z}_{\omega}^{\dag} & \mathrm{0}\end{pmatrix}\left(\begin{array}{cc}
2\mathrm{i}\tan\frac{\omega}{2} & \frac{-2\mathrm{i}e^{\mathrm{i}\omega}}{1+e^{\mathrm{i}\omega}}\\
\frac{-2\mathrm{i}}{1+e^{\mathrm{i}\omega}} & \frac{-1}{1+e^{\mathrm{i}\omega}}
\end{array}\right)\begin{pmatrix}\mathbf{z}_{\omega}\\
0
\end{pmatrix}\nonumber \\
 & =2\mathrm{i}\sum_{\omega}\mathbf{z}_{\omega}^{\dag}\mathbf{z}_{\omega}\tan\frac{\omega}{2}.
\end{align}

Finally, the discrete thermal coherent-state path integral in the
Weyl order is
\begin{equation}
\mathcal{Z}=\underset{N\text{ odd}}{\lim_{N\rightarrow\infty}}2^{\left(N-1\right)M}\int\left[\mathcal{D}\mathbf{z}^{\dag}\mathcal{D}\mathbf{z}\right]\exp\left[2\mathrm{i}\sum_{\omega}\mathbf{z}_{\omega}^{\dag}\mathbf{z}_{\omega}\tan\frac{\omega}{2}-\sum_{l=0}^{N-1}\Delta\mathcal{H}\left(\mathbf{z}_{l}^{\dag},\mathbf{z}_{l}\right)\right].\label{discrete Weyl}
\end{equation}
Similarly to the construction based on the anti-normal order, the
Hamiltonian takes two arguments at the same imaginary time moment.
However, the Berry's phase term (the one in the exponential not involving
the Hamiltonian) is very different. It is easy to check, that it behaves
in the same way as its previous versions for low frequencies, but
it differs significantly for higher. Therefore again, the continuum
version of Eq. (\ref{discrete Weyl}) has the same structure as earlier
constructions, but with a different Hamiltonian form.

The peculiar restriction regarding parity of $N$ deserves a comment.
For even $N$ frequency $\omega=\pi$ is allowed (i. e. it enters
the summation $\sum_{\omega}$). Then formally matrix $\mathbf{M}$
becomes singular and the performed construction cannot be finished.
From a less formal viewpoint, $\tan\omega/2$ blows up to $\pm\infty$
at $\omega=\pi$. This suppresses any fluctuations of $\mathbf{z}_{\omega=\pi}$
keeping it zero. The integral in Eq. (\ref{discrete Weyl}) becomes
zero, but its prefactor becomes infinite (since $\det\left(-\mathbf{M}\right)=0$).
It suggests that the presented construction can be done for even $N$,
if $\mathbf{z}_{\omega=\pi}=0$ is somehow enforced. However, this
effort is not necessary, since restricting to odd $N$'s does the
job.

\section{Tests on the harmonic oscillator}\label{sec:Tests-on-the}

Three different discrete constructions of the thermal coherent-state
path integral have been derived. They are all exact, but they produce
three different continuum limits, from which at most one can be correct.
A natural strategy is to test the continuum path integral with different
orderings of the Hamiltonian for the simplest model, which is a single
mode harmonic oscillator. Additionally, the normal-ordered discrete
construction should be evaluated, because it throws some light on
the subtleties involved (pointed out also in \citep{bergeron1992coherent})
and serves as a good double-check.

The Hamiltonian reads
\begin{equation}
\hat{H}=Aa^{\dag}a\label{SHOH}
\end{equation}
and its exact solution is
\begin{equation}
\mathcal{Z}_{0}=\sum_{n=0}^{\infty}e^{-\beta An}=\frac{1}{1-e^{-\beta A}}.
\end{equation}
For further technical reasons, it is easier to find a derivative of
the free energy with respect to $A$, which is
\begin{equation}
\frac{\partial\mathcal{F}}{\partial A}=\frac{1}{2}\left(\coth\frac{\beta A}{2}-1\right).\label{dF/dA corr}
\end{equation}
The derivative removes any additive constant terms from $\mathcal{F}$,
so any spotted discrepancy would be more than just an irrelevant shift.

\subsection{Discrete normal-ordered version}

Inserting Hamiltonian (\ref{SHOH}) into Eq. (\ref{Z normal}) gives
\begin{equation}
\mathcal{Z}=\lim_{N\rightarrow\infty}\int\left[\mathcal{D}\bar{z}\mathcal{D}z\right]\exp\left\{ -\sum_{l=0}^{N-1}\left[\bar{z}_{l}\left(z_{l}-z_{l+1}\right)+\frac{\beta A}{N}\bar{z}_{l}z_{l+1}\right]\right\} ,
\end{equation}
which is a complex Gaussian integral:
\begin{equation}
\mathcal{Z}=\lim_{N\rightarrow\infty}\int\left[\mathcal{D}\bar{z}\mathcal{D}z\right]\exp\left[-\sum_{ll^{\prime}}\bar{z}_{l}\underbrace{\left(\delta_{ll^{\prime}}-\delta_{l+1,l^{\prime}}+\frac{\beta A}{N}\delta_{l+1,l^{\prime}}\right)}_{\mathbf{G}}z_{l^{\prime}}\right].
\end{equation}
Eigenvalues of the matrix in round brackets are $1-e^{\mathrm{i}\omega}+\frac{\beta A}{N}e^{\mathrm{i}\omega}$
(which is apparent after switching to the frequency domain defined
by Eq. (\ref{fourier})). This allows to calculate the free energy:
\begin{align}
\mathcal{F} & =\frac{1}{\beta}\mathrm{Tr}\ln\mathbf{G}\nonumber \\
 & =\frac{1}{\beta}\sum_{\omega}\ln\left(1-e^{\mathrm{i}\omega}+\frac{\beta A}{N}e^{\mathrm{i}\omega}\right)
\end{align}
and its derivative:
\begin{align}
 & \frac{\partial\mathcal{F}}{\partial A}=\sum_{\omega}\frac{1}{N\left(e^{-\mathrm{i}\omega}-1+\frac{\beta A}{N}\right)}\nonumber \\
 & =\sum_{n=-N_{1}}^{N_{1}}\frac{1}{N\left(e^{-2\pi\mathrm{i}n/N}-1+\frac{\beta A}{N}\right)}+\frac{1}{N}\underset{n\text{s}}{\sum_{\text{remaining}}}\frac{1}{e^{-2\pi\mathrm{i}n/N}-1+\frac{\beta A}{N}}\nonumber \\
 & =\sum_{n=-N_{1}}^{N_{1}}\frac{1}{\beta A+N\left[-2\pi\mathrm{i}\frac{n}{N}+\mathcal{O}\left(\frac{n^{2}}{N^{2}}\right)\right]}+\frac{1}{N}\underset{n\text{s}}{\sum_{\text{remaining}}}\frac{1+\mathcal{O}\left(N_{1}^{-1}\right)}{e^{-2\pi\mathrm{i}n/N}-1}\nonumber \\
 & =\sum_{n=-N_{1}}^{N_{1}}\frac{1+\mathcal{O}\left(n/N\right)}{\beta A-2\pi\mathrm{i}n}+\int_{\left(-1/2,1/2\right)\backslash\left(-N_{1}/N,N_{1}/N\right)}\frac{\mathrm{d}x}{e^{-2\pi\mathrm{i}x}-1}+\mathcal{O}\left(N_{1}^{-1}\right)\nonumber \\
 & \overset{1\ll N_{1}\ll N}{\longrightarrow}\mathrm{p.v.}\sum_{n=-\infty}^{\infty}\frac{1}{\beta A-2\pi\mathrm{i}n}+\mathrm{p.v.}\int_{-1/2}^{1/2}\frac{\mathrm{d}x}{e^{-2\pi\mathrm{i}x}-1}\nonumber \\
 & =\frac{1}{2}\coth\left(\frac{1}{2}\beta A\right)-\frac{1}{2\pi\mathrm{i}}\mathrm{p.v.}\oint\frac{\mathrm{d}z}{z-1}\nonumber \\
 & =\frac{1}{2}\left(\coth\frac{\beta A}{2}-1\right).
\end{align}
The correct result from Eq. (\ref{dF/dA corr}) is recovered, but
a rather intricate limit had to be taken. Instead of a sum just turning
into an integral, as it usually happens in similar limits, a series
and an integral emerged (both marked as their principal values). The
first object comes from lower frequencies and the second from higher
ones.

\subsection{Continuum version}

Rewriting Eq. (\ref{cont conj}) for the harmonic oscillator yields
\begin{equation}
\mathcal{Z}=\int\left[\mathcal{D}\bar{z}\mathcal{D}z\right]\exp\left\{ -\int_{0}^{\beta}\mathrm{d}\tau\left[\bar{z}\left(\tau\right)\frac{\partial}{\partial\tau}z\left(\tau\right)+A\bar{z}\left(\tau\right)z\left(\tau\right)\right]\right\} ,
\end{equation}
which is a functional complex Gaussian integral:
\begin{equation}
\mathcal{Z}=\int\left[\mathcal{D}\bar{z}\mathcal{D}z\right]\exp\left[-\int_{0}^{\beta}\mathrm{d}\tau\,\bar{z}\left(\tau\right)\underbrace{\left(\frac{\partial}{\partial\tau}+A\right)}_{\hat{G}}z\left(\tau\right)\right].
\end{equation}

If we write a Fourier series for the path $z\left(\tau\right)$, Matsubara
frequency domain can be introduced:
\begin{equation}
z\left(\tau\right)=\frac{1}{\sqrt{\beta}}\sum_{\ell=-\infty}^{\infty}z_{\ell}e^{\mathrm{i}\omega_{\ell}\tau},\;\omega_{\ell}=\frac{2\pi}{\beta}\ell.\label{cont fourier}
\end{equation}
It should be noted that here, in the continuous case, frequencies
are differently defined than in the discrete case (Eq. (\ref{fourier})).
In this domain, operator $\hat{G}$ becomes simply
\begin{equation}
\hat{G}=\mathrm{i}\omega_{\ell}+A,
\end{equation}
which leads to
\begin{equation}
\mathcal{F}=\frac{1}{\beta}\sum_{\ell}\ln\left(\mathrm{i}\omega_{\ell}+A\right)
\end{equation}
and
\begin{align}
\frac{\partial\mathcal{F}}{\partial A} & =\sum_{\ell}\frac{1}{\mathrm{i}\beta\omega_{\ell}+\beta A}\nonumber \\
 & =\frac{1}{2}\coth\frac{\beta A}{2}\;\text{\lightning}.\label{wrong}
\end{align}

The result is very similar to the exact one, but it differs by a term
identified in the previous subsection as the contribution from high
frequencies. The reason for this can be easily understood. Namely,
in the continuum limit, although infinitely many frequencies are available,
they correspond only to the low-frequency part of the discrete version,
as the other part disappears from the view after introducing continuous
functions.

The final answer can be easily modified for other orderings of the
Hamiltonian. Since
\begin{equation}
\hat{H}=Aa^{\dag}a=Aaa^{\dag}-A=\frac{aa^{\dag}+a^{\dag}a}{2}-\frac{A}{2},
\end{equation}
we have
\begin{equation}
h\left(\bar{z},z\right)=H\left(\bar{z},z\right)-A
\end{equation}
and
\begin{equation}
\mathcal{H}\left(\bar{z},z\right)=H\left(\bar{z},z\right)-A/2.
\end{equation}
Now this is apparent that replacing the normal-ordered Hamiltonian
by its Weyl-ordered form in the continuum path integral gives the
correct result.

Although functional integration was not explicitly defined here it
requires the following formalization. The integration is performed
over Fourier amplitudes from Eq. (\ref{cont fourier}) up to some
absolute cut-off frequency, which finally is taken to grow to infinity.
Thus the series in Eq. (\ref{wrong}) should be understood as its
principal value, similarly to the discrete case.

\section{Identifying and understanding the correct path integral}\label{sec:Identifying-and-understanding}

Example of the single-mode harmonic oscillator indicated the Weyl-ordered
path integral as the best candidate for constructing a correct continuum
version. Indeed, this is also what Kordas et al. \citep{kordas2014coherent}
have found, additionally showing that even for a single-site Bose-Hubbard
Hamiltonian Weyl-order is well-suited. However, they did not provide
any justification for suitability of the construction for every polynomial
Hamiltonian (although it has been claimed). The aim of this section
is to provide such a general argument and present a formal construction
of a continuum thermal coherent-state path integral.

The main idea of the reasoning is to start with the exact discrete
Weyl-ordered path integral (Eq. (\ref{discrete Weyl})) and integrate
out high frequencies. Such an operation constitutes a kind of imaginary
time renormalization (in the frequency domain) and the crucial part
is to show that the Hamiltonian flow is not present, i. e. the Hamiltonian
remains invariant.

First, we note that Fourier components $\mathbf{z}_{\omega}$ of the
discrete paths are related to $\mathbf{z}_{l}$ by a unitary transform,
so the complex integration can be equivalently carried over $\mathbf{z}_{\omega}$:
\begin{equation}
\int\left[\mathcal{D}\mathbf{z}^{\dag}\mathcal{D}\mathbf{z}\right]\,\left(\cdots\right)=\int\left(\prod_{n}\mathrm{d}\mathbf{z}_{\omega_{n}}^{\dag}\mathrm{d}\mathbf{z}_{\omega_{n}}\right)\,\left(\cdots\right).
\end{equation}
Since $N$ is odd, it can be written as $N=2B+1$ for some large integer
$B$. Nonzero frequencies $\omega_{n}=2\pi n/N$ ($n\in\left\{ -B,\dots,B\right\} $)
can be paired with their opposite values (i. e. $\omega_{n}$ with
$\omega_{-n}$) and corresponding Fourier components in a pair should
be always integrated out together. Suppose we have already integrated
out all $\omega_{n}$'s for $\left|n\right|>B^{\prime}$, so that
the partition function obtained is $\mathcal{Z}_{B,B^{\prime}}$ and
the renormalized Hamiltonian is $\mathcal{H}_{B,B^{\prime}}$. Now,
we want to integrate over $\mathbf{z}_{\pm\omega_{B^{\prime}}}^{\dag},\mathbf{z}_{\pm\omega_{B^{\prime}}}$
to determine the renormalization group transformation. We begin by
visually splitting integration over these variables from the rest:
\begin{align}
 & \mathcal{Z}_{B,B^{\prime}}=c_{B,B^{\prime}}\int\left(\prod_{\left|n\right|<B^{\prime}}\mathrm{d}\mathbf{z}_{\omega_{n}}^{\dag}\mathrm{d}\mathbf{z}_{\omega_{n}}\right)\exp\left(2\mathrm{i}\sum_{\left|n\right|<B^{\prime}}\mathbf{z}_{\omega_{n}}^{\dag}\mathbf{z}_{\omega_{n}}\tan\frac{\omega_{n}}{2}\right)\nonumber \\
 & \times\int\mathrm{d}\mathbf{z}_{\omega_{B^{\prime}}}^{\dag}\mathrm{d}\mathbf{z}_{\omega_{B^{\prime}}}\mathrm{d}\mathbf{z}_{-\omega_{B^{\prime}}}^{\dag}\mathrm{d}\mathbf{z}_{-\omega_{B^{\prime}}}\exp\left[2\mathrm{i}\left(\mathbf{z}_{\omega_{B^{\prime}}}^{\dag}\mathbf{z}_{\omega_{B^{\prime}}}-\mathbf{z}_{-\omega_{B^{\prime}}}^{\dag}\mathbf{z}_{-\omega_{B^{\prime}}}\right)\tan\frac{\omega_{B^{\prime}}}{2}\right]\nonumber \\
 & \times\exp\left[-\sum_{l=0}^{N-1}\Delta\mathcal{H}_{B,B^{\prime}}\left(\mathbf{z}_{l}^{\dag},\mathbf{z}_{l}\right)\right].\label{renorm}
\end{align}

The procedure will be repeated until $B^{\prime}$ reaches its minimal
value denoted as $b$. Finally, a double limit $1\ll b\ll B$ should
be taken. In this sense, only high-frequencies are removed. If we
focus on a single-frequency contribution to the Berry's phase, namely
$2\mathrm{i}\mathbf{z}_{\omega_{n}}^{\dag}\mathbf{z}_{\omega_{n}}\tan\frac{\omega_{n}}{2}$,
it can be seen that it increases with frequency in magnitude. The
coefficient involved can be estimated as $2\left|\tan\left(\omega_{n}/2\right)\right|\geq2\pi n/N$.
Thus for high frequencies, this terms dominates the Hamiltonian, which
can be treated as a perturbation. Moreover, due to the rapid phase
oscillations introduced by the Berry's phase factor, magnitude of
$\mathbf{z}_{\omega_{n}}$, effectively contributing to the integral,
is on the order of $1/\sqrt{2\tan\frac{\omega_{n}}{2}}$. This falls
to zero for large $n$, but not quickly enough to make the paths obviously
smooth.

Therefore, it remains to expand the Hamiltonian in a power series
with respect to $\mathbf{z}_{\pm\omega_{B^{\prime}}}^{\dag},\mathbf{z}_{\pm\omega_{B^{\prime}}}$
and perform a Gaussian integral. For this purpose we notice that
\begin{align}
\frac{\partial}{\partial z_{\omega_{n},m}}\sum_{l=0}^{N-1}\Delta\mathcal{H}\left(\mathbf{z}_{l}^{\dag},\mathbf{z}_{l}\right) & =\sum_{l^{\prime}=0}^{N-1}\frac{\partial z_{l^{\prime},m}}{\partial z_{\omega_{n},m}}\frac{\partial}{\partial z_{l^{\prime},m}}\sum_{l=0}^{N-1}\Delta\mathcal{H}\left(\mathbf{z}_{l}^{\dag},\mathbf{z}_{l}\right)\nonumber \\
 & =\Delta\sum_{l^{\prime}=0}^{N-1}\frac{\partial z_{l^{\prime},m}}{\partial z_{\omega_{n},m}}\frac{\partial\mathcal{H}}{\partial z_{l^{\prime},m}}\left(\mathbf{z}_{l^{\prime}}^{\dag},\mathbf{z}_{l^{\prime}}\right)\nonumber \\
 & =\Delta\frac{1}{\sqrt{N}}\sum_{l=0}^{N-1}\frac{\partial\mathcal{H}\left(\mathbf{z}_{l}^{\dag},\mathbf{z}_{l}\right)}{\partial z_{l,m}}e^{\mathrm{i}\omega_{n}l}\label{der1}
\end{align}
and analogously
\begin{equation}
\frac{\partial}{\partial\bar{z}_{\omega_{n},m}}\sum_{l=0}^{N-1}\Delta\mathcal{H}\left(\mathbf{z}_{l}^{\dag},\mathbf{z}_{l}\right)=\Delta\frac{1}{\sqrt{N}}\sum_{l=0}^{N-1}\frac{\partial\mathcal{H}\left(\mathbf{z}_{l}^{\dag},\mathbf{z}_{l}\right)}{\partial\bar{z}_{l,m}}e^{-\mathrm{i}\omega_{n}l}.\label{der2}
\end{equation}
From Eqs. (\ref{der1}) and (\ref{der2}) one can see that taking
a derivative of the sum over the Hamiltonian brings a factor of $N^{-1/2}$
and locally modifies the Hamiltonian by $e^{\mathrm{i}\omega_{n}l}\partial/\partial z_{l,m}$
or $e^{-\mathrm{i}\omega_{n}l}\partial/\partial\bar{z}_{l,m}$ preserving
the structure of the sum. Thus:
\begin{equation}
\frac{\partial}{\partial\bar{z}_{\omega_{n},m}\partial z_{\omega_{n},m}}\sum_{l=0}^{N-1}\Delta\mathcal{H}\left(\mathbf{z}_{l}^{\dag},\mathbf{z}_{l}\right)=\Delta\frac{1}{N}\sum_{l=0}^{N-1}\frac{\partial^{2}\mathcal{H}\left(\mathbf{z}_{l}^{\dag},\mathbf{z}_{l}\right)}{\partial\bar{z}_{l,m}\partial z_{l,m}}
\end{equation}
and higher derivatives can be easily constructed. Using the chain
rule and Eqs. (\ref{der1}) and (\ref{der2}), derivative of the exponential
can be obtained:
\begin{align}
 & \frac{\partial^{2}}{\partial\bar{z}_{\pm\omega_{B^{\prime}},m}\partial z_{\pm\omega_{B^{\prime}},m}}\exp\left[-\sum_{l=0}^{N-1}\Delta\mathcal{H}\left(\mathbf{z}_{l}^{\dag},\mathbf{z}_{l}\right)\right]=\nonumber \\
 & =\frac{1}{N}\exp\left[-\sum_{l=0}^{N-1}\Delta\mathcal{H}\left(\mathbf{z}_{l}^{\dag},\mathbf{z}_{l}\right)\right]\nonumber \\
 & \times\left[\left|\sum_{l=0}^{N-1}\Delta\frac{\partial\mathcal{H}\left(\mathbf{z}_{l}^{\dag},\mathbf{z}_{l}\right)}{\partial z_{l,m}}e^{\pm\mathrm{i}\omega_{B^{\prime}}l}\right|^{2}-\sum_{l=0}^{N-1}\Delta\frac{\partial^{2}\mathcal{H}\left(\mathbf{z}_{l}^{\dag},\mathbf{z}_{l}\right)}{\partial\bar{z}_{l,m}\partial z_{l,m}}\right].\label{double der}
\end{align}
Only derivatives composed of $\partial^{2}/\left(\partial\bar{z}_{\pm\omega_{B^{\prime}},m}\partial z_{\pm\omega_{B^{\prime}},m}\right)$
lead to nonzero terms in the integration from Eq. (\ref{renorm})
and each additional double derivative brings at least a factor of
$1/N$. The zeroth order term (in expanding the exponential) can be
simply written as
\begin{equation}
\exp\left[-\sum_{l=0}^{N-1}\Delta\mathcal{H}\left(\mathbf{\tilde{z}}_{l}^{\dag},\mathbf{\tilde{z}}_{l}\right)\right],
\end{equation}
where the tilde means removing the frequencies $\pm\omega_{B^{\prime}}$
from the Fourier expansion of $\mathbf{z}_{l}$ (all higher frequencies
are already not present).

The expected Gaussian integral arising from Eq. (\ref{renorm}) becomes:
\begin{align}
 & \mathcal{Z}_{B,B^{\prime}}=\frac{c_{B,B^{\prime}}}{\left(4\tan^{2}\frac{\omega_{B^{\prime}}}{2}\right)^{M}}\int\left(\prod_{\left|n\right|<B^{\prime}}\mathrm{d}\mathbf{z}_{\omega_{n}}^{\dag}\mathrm{d}\mathbf{z}_{\omega_{n}}\right)\exp\left(2\mathrm{i}\sum_{\left|n\right|<B^{\prime}}\mathbf{z}_{\omega_{n}}^{\dag}\mathbf{z}_{\omega_{n}}\tan\frac{\omega_{n}}{2}\right)\nonumber \\
 & \times\left\{ \exp\left[-\sum_{l=0}^{N-1}\Delta\mathcal{H}_{B,B^{\prime}}\left(\mathbf{\tilde{z}}_{l}^{\dag},\mathbf{\tilde{z}}_{l}\right)\right]\right.\nonumber \\
 & \left.+\frac{\mathrm{i}}{2\tan\frac{\omega_{B^{\prime}}}{2}}\sum_{\pm,m}\frac{\pm\partial^{2}\exp\left[-\sum_{l=0}^{N-1}\Delta\mathcal{H}_{B,B^{\prime}}\left(\mathbf{z}_{l}^{\dag},\mathbf{z}_{l}\right)\right]}{\partial\bar{z}_{\pm\omega_{B^{\prime}},m}\partial z_{\pm\omega_{B^{\prime}},m}}+\mathcal{O}\left(\frac{1}{N^{2}}\frac{1}{\tan^{2}\frac{\omega_{B^{\prime}}}{2}}\right)\right\} .\label{Gaussian}
\end{align}
Since tangent is an odd function, opposite frequencies have opposite
contributions to the leading term (summation over the $\pm$ sign).
The second term in square brackets in Eq. (\ref{double der}) cancels
out during this summation. Focusing on the term that survives, substituting
it into Eq. (\ref{Gaussian}) and rearranging leads to
\begin{align}
 & \mathcal{Z}_{B,B^{\prime}}=\frac{c_{B,B^{\prime}}}{\left(4\tan^{2}\frac{\omega_{B^{\prime}}}{2}\right)^{M}}\int\left(\prod_{\left|n\right|<B^{\prime}}\mathrm{d}\mathbf{z}_{\omega_{n}}^{\dag}\mathrm{d}\mathbf{z}_{\omega_{n}}\right)\exp\left(2\mathrm{i}\sum_{\left|n\right|<B^{\prime}}\mathbf{z}_{\omega_{n}}^{\dag}\mathbf{z}_{\omega_{n}}\tan\frac{\omega_{n}}{2}\right)\nonumber \\
 & \times\exp\left[-\sum_{l=0}^{N-1}\Delta\mathcal{H}_{B,B^{\prime}}\left(\mathbf{\tilde{z}}_{l}^{\dag},\mathbf{\tilde{z}}_{l}\right)\right]\left[1+\sum_{\pm,m}\frac{\pm\mathrm{i}\beta^{2}}{2N^{2}\tan\frac{\omega_{B^{\prime}}}{2}}\right.\nonumber \\
 & \left.\times\left|\frac{1}{\sqrt{N}}\sum_{l=0}^{N-1}\frac{\partial\mathcal{H}_{B,B^{\prime}}}{\partial z_{l,m}}\left(\mathbf{\tilde{z}}_{l}^{\dag},\mathbf{\tilde{z}}_{l}\right)e^{\pm\mathrm{i}\omega_{B^{\prime}}l}\right|^{2}+\mathcal{O}\left(\frac{1}{N^{2}}\frac{1}{\tan^{2}\frac{\omega_{B^{\prime}}}{2}}\right)\right].\label{Gaussian-1}
\end{align}
First expression in the third line of Eq. (\ref{Gaussian-1}) is a
magnitude squared of a single Fourier component (at frequency $\pm\omega_{B^{\prime}}$)
of a sequence $\left(\partial/\partial z_{l,m}\right)\mathcal{H}_{B,B^{\prime}}\left(\mathbf{\tilde{z}}_{l}^{\dag},\mathbf{\tilde{z}}_{l}\right)$.
Although sequence $\mathbf{\tilde{z}}_{l}$ does not contain such
a component, Hamiltonian as a nonlinear function can give rise to
its nonzero value. Still, magnitudes of such components should fall
in the same asymptotic pace with $\omega$ for both $\mathbf{\tilde{z}}_{l}$
and $\left(\partial/\partial z_{l,m}\right)\mathcal{H}_{B,B^{\prime}}\left(\mathbf{\tilde{z}}_{l}^{\dag},\mathbf{\tilde{z}}_{l}\right)$.
Therefore, the absolute value term (including the square) can be classified
as $\mathcal{O}\left(1/\tan\frac{\omega_{n}}{2}\right)$. Since $N^{2}\tan^{2}\frac{\omega_{B^{\prime}}}{2}>\frac{1}{4}N^{2}\omega_{B^{\prime}}^{2}\sim B^{\prime2}$,
the entire curly bracket in Eq. (\ref{Gaussian-1}) can be written
as $1+\mathcal{O}\left(B^{\prime-2}\right)=\exp\left[\mathcal{O}\left(B^{\prime-2}\right)\right]$,
which leads to
\begin{align}
 & \mathcal{Z}_{B,B^{\prime}}=\frac{c_{B,B^{\prime}}}{\left(4\tan^{2}\frac{\omega_{B^{\prime}}}{2}\right)^{M}}\int\left(\prod_{\left|n\right|<B^{\prime}}\mathrm{d}\mathbf{z}_{\omega_{n}}^{\dag}\mathrm{d}\mathbf{z}_{\omega_{n}}\right)\exp\left(2\mathrm{i}\sum_{\left|n\right|<B^{\prime}}\mathbf{z}_{\omega_{n}}^{\dag}\mathbf{z}_{\omega_{n}}\tan\frac{\omega_{n}}{2}\right)\nonumber \\
 & \times\exp\left\{ -\sum_{l=0}^{N-1}\Delta\left[\mathcal{H}_{B,B^{\prime}}\left(\mathbf{\tilde{z}}_{l}^{\dag},\mathbf{\tilde{z}}_{l}\right)+\mathcal{O}\left(B^{\prime-2}\right)\right]\right\} .\label{renormalized}
\end{align}
Of course, renormalization does not change the partition function,
so $\mathcal{Z}_{B,B^{\prime}}$ does not depend on $B^{\prime}$.
We can now read off renormalized Hamiltonian and normalizing factor
from Eq. (\ref{renormalized}):
\begin{equation}
\begin{cases}
c_{B,B^{\prime}-1}=\left(4\tan^{2}\frac{\omega_{B^{\prime}}}{2}\right)^{-M}c_{B,B^{\prime}}\\
\mathcal{H}_{B,B^{\prime}-1}=\mathcal{H}_{B,B^{\prime}}+\mathcal{O}\left(B^{\prime-2}\right)
\end{cases}.\label{group flow}
\end{equation}

The procedure is repeated for $B^{\prime}$ starting from $B$ and
decreasing in steps of $1$ down to $b$. Thus the accumulated correction
to the Hamiltonian is on the order of
\begin{equation}
\sum_{B^{\prime}=b+1}^{B}\mathcal{O}\left(B^{\prime-2}\right)\overset{1\ll b\ll B}{\longrightarrow}0.\label{main trick}
\end{equation}
Regime $1\ll b\ll B$ means formally a double limit $b\rightarrow\infty$
and $B/b\rightarrow\infty$. To show that Eq. (\ref{main trick})
holds, we start by noting that a series $\sum_{B^{\prime}=1}^{\infty}1/B^{\prime2}$
is convergent. Then, on the basis of Cauchy's convergence test,
\begin{equation}
\sum_{B^{\prime}=b+1}^{B}B^{\prime-2}\overset{1\ll b\ll B}{\longrightarrow}0.\label{simple}
\end{equation}
Next, $\mathcal{O}\left(B^{\prime-2}\right)$, by definition, can
be dominated by some $\mathrm{const.}/B^{\prime2}$, which together
with Eq. (\ref{simple}) implies Eq. (\ref{main trick}).

After tracing out all high frequencies, we arrive at
\begin{align}
 & \mathcal{Z}_{B,b}=c_{B,b}\int\left(\prod_{\left|n\right|\leq b}\mathrm{d}\mathbf{z}_{\omega_{n}}^{\dag}\mathrm{d}\mathbf{z}_{\omega_{n}}\right)\exp\left(2\mathrm{i}\sum_{\left|n\right|\leq b}\mathbf{z}_{\omega_{n}}^{\dag}\mathbf{z}_{\omega_{n}}\tan\frac{\omega_{n}}{2}\right)\nonumber \\
 & \times\exp\left[-\sum_{l=0}^{N-1}\Delta\mathcal{H}\left(\mathbf{z}_{l}^{\dag},\mathbf{z}_{l}\right)\right].\label{B,b}
\end{align}
The Hamiltonian is written without subscripts, because it is essentially
unchanged after the renormalization. Since the highest frequency present
in $\mathbf{z}_{l}$ is $\omega_{b}$, $\mathcal{H}\left(\mathbf{z}_{l}^{\dag},\mathbf{z}_{l}\right)$
is practically smooth compared to the dense sampling in the sum $\sum_{l=0}^{N-1}\Delta\mathcal{H}$.
Therefore, it can be safely replaced by an integral. Also $2\tan\frac{\omega_{n}}{2}\cong\omega_{n}$
(due to $b\ll N$). To tailor the discrete and a naturally corresponding
continuous representation of the paths (which now exists), the following
Fourier transforms must be matched:
\begin{equation}
\begin{cases}
\mathbf{z}_{l}=\frac{1}{\sqrt{N}}\sum_{\omega}\mathbf{z}_{\omega}e^{\mathrm{i}\omega l}, & \omega_{n}=\frac{2\pi}{N}n,\left|n\right|\leq b\\
\mathbf{z}\left(\tau\right)=\frac{1}{\sqrt{\beta}}\sum_{\ell=-b}^{b}\mathbf{z}_{\ell}^{\mathrm{cont}}e^{\mathrm{i}\omega_{\ell}^{\mathrm{cont}}\tau}, & \omega_{\ell}^{\mathrm{cont}}=\frac{2\pi}{\beta}\ell
\end{cases},
\end{equation}
to satisfy $\mathbf{z}_{l}=\mathbf{z}\left(\Delta l\right)$ for every
$l$. This is provided by
\begin{equation}
\mathbf{z}_{\omega_{n}}=\sqrt{\frac{N}{\beta}}\mathbf{z}_{n}^{\mathrm{cont}}.
\end{equation}
Changing variables from $\mathbf{z}_{\omega_{n}}$ to $\mathbf{z}_{n}^{\mathrm{cont}}$
and realizing the mentioned limit transitions turns Eq. (\ref{B,b})
into
\begin{align}
 & \mathcal{Z}_{B,b}=c_{B,b}\int\left[\prod_{\left|n\right|\leq b}\left(\frac{N}{\beta}\right)^{M}\mathrm{d}\mathbf{z}_{n}^{\mathrm{cont}\dag}\mathrm{d}\mathbf{z}_{n}^{\mathrm{cont}}\right]\exp\left(\sum_{\left|n\right|\leq b}\mathbf{z}_{n}^{\mathrm{cont}\dag}\mathrm{i}\omega_{n}^{\mathrm{cont}}\mathbf{z}_{n}^{\mathrm{cont}}\right)\nonumber \\
 & \times\exp\left[-\int_{0}^{\beta}\mathrm{d}\tau\,\mathcal{H}\left(\mathbf{z}^{\dag}\left(\tau\right),\mathbf{z}\left(\tau\right)\right)\right].\label{B,b-1}
\end{align}
Replacing the Berry's phase by a corresponding integral expression
gives
\begin{align}
 & \mathcal{Z}_{B,b}=\left(\frac{N}{\beta}\right)^{M\left(2b+1\right)}c_{B,b}\int\left(\prod_{\left|n\right|\leq b}\mathrm{d}\mathbf{z}_{n}^{\mathrm{cont}\dag}\mathrm{d}\mathbf{z}_{n}^{\mathrm{cont}}\right)\exp\left[\int_{0}^{\beta}\mathrm{d}\tau\,\mathbf{z}^{\dagger}\left(\tau\right)\frac{\partial}{\partial\tau}\mathbf{z}\left(\tau\right)\right]\nonumber \\
 & \times\exp\left[-\int_{0}^{\beta}\mathrm{d}\tau\,\mathcal{H}\left(\mathbf{z}^{\dag}\left(\tau\right),\mathbf{z}\left(\tau\right)\right)\right].\label{main result}
\end{align}
Equation (\ref{main result}) is the anticipated continuum path integral
derived from an exact discrete construction.

It is possible to obtain a precise expression for the normalization
prefactor in Eq. (\ref{main result}). Details of the calculation
are given in Appendix \ref{sec:Calculation-of-the}. Including it,
the constructed integral becomes:
\begin{align}
 & \mathcal{Z}=\lim_{b\rightarrow\infty}\left[\beta^{-\left(2b+1\right)}\left(2\pi\right)^{2b}\left(b!\right)^{2}\right]^{M}\int\left(\prod_{\left|n\right|\leq b}\mathrm{d}\mathbf{z}_{n}^{\mathrm{cont}\dag}\mathrm{d}\mathbf{z}_{n}^{\mathrm{cont}}\right)\nonumber \\
 & \times\exp\left[\int_{0}^{\beta}\mathrm{d}\tau\,\mathbf{z}^{\dagger}\left(\tau\right)\frac{\partial}{\partial\tau}\mathbf{z}\left(\tau\right)-\int_{0}^{\beta}\mathrm{d}\tau\,\mathcal{H}\left(\mathbf{z}^{\dag}\left(\tau\right),\mathbf{z}\left(\tau\right)\right)\right],\label{main result-1}
\end{align}
which can be complex conjugated (since $\mathcal{Z}$ is real) to
give
\begin{align}
 & \mathcal{Z}=\lim_{b\rightarrow\infty}\left[\beta^{-\left(2b+1\right)}\left(2\pi\right)^{2b}\left(b!\right)^{2}\right]^{M}\int\left(\prod_{\left|n\right|\leq b}\mathrm{d}\mathbf{z}_{n}^{\mathrm{cont}\dag}\mathrm{d}\mathbf{z}_{n}^{\mathrm{cont}}\right)\nonumber \\
 & \times\exp\left[-\int_{0}^{\beta}\mathrm{d}\tau\,\mathbf{z}^{\dagger}\left(\tau\right)\frac{\partial}{\partial\tau}\mathbf{z}\left(\tau\right)-\int_{0}^{\beta}\mathrm{d}\tau\,\mathcal{H}\left(\mathbf{z}^{\dag}\left(\tau\right),\mathbf{z}\left(\tau\right)\right)\right].\label{main result-2}
\end{align}

\section{Comparison with the construction of Klauder}\label{sec:Klauder}

The construction derived by Klauder and Daubechies \citep{daubechies1985quantum,klauder2010modern}
concerns real-time propagator
\begin{align}
 & \left\langle \mathbf{p}^{\prime\prime},\mathbf{q}^{\prime\prime}\right|e^{-\mathrm{i}T\hat{H}/\hslash}\left|\mathbf{p}^{\prime},\mathbf{q}^{\prime}\right\rangle =\nonumber \\
 & =\lim_{\nu\rightarrow\infty}\left(2\pi\hslash\right)^{M}e^{M\nu T/2\hslash}\int e^{\left(\mathrm{i}/\hslash\right)\int\left[\mathbf{p}^{T}\mathrm{d}\mathbf{q}-h\left(\mathbf{p},\mathbf{q}\right)\mathrm{d}t\right]}\mathrm{d}\mu_{W}^{\nu}\left(\mathbf{p},\mathbf{q}\right).\label{real-time-Klaud}
\end{align}
$\left|\mathbf{p},\mathbf{q}\right\rangle $ stands for a coherent
state with momentum $\mathbf{p}$ and position $\mathbf{q}$ (each
having $M$ components). $h\left(\mathbf{p},\mathbf{q}\right)$ is
the $P$-representation of the Hamiltonian $\hat{H}$ as in Eq. (\ref{P-rep}).
$\int\mathbf{p}^{T}\mathrm{d}\mathbf{q}$ has to be understood as
the Stratonovich integral ($T$ stands for transpose) and $\mathrm{d}\mu_{W}^{\nu}\left(p,q\right)$
stands for the Wiener measure (independent for all components) with
diffusion constant $\nu$. Endpoints of the trajectories are fixed
at $\left(\mathbf{p}^{\prime\prime},\mathbf{q}^{\prime\prime}\right)$
and $\left(\mathbf{p}^{\prime},\mathbf{q}^{\prime}\right)$. Assuming
that a natural transition to the imaginary time can be done, and switching
to
\begin{equation}
\begin{cases}
\mathbf{z}=\left(\mathbf{q}+\mathrm{i}\mathbf{p}\right)/\sqrt{2}\\
\mathbf{z}^{\dag}=\left(\mathbf{q}^{T}-\mathrm{i}\mathbf{p}^{T}\right)/\sqrt{2}
\end{cases},
\end{equation}
leads to an expression for the partition function (where the trajectories
are periodic and starting at $\mathbf{z}^{\prime}$):
\begin{align}
\mathcal{Z} & =\int\mathrm{d}\mathbf{z}^{\prime\dag}\mathrm{d}\mathbf{z}^{\prime}\lim_{\nu\rightarrow\infty}\left(2\pi e^{-\mathrm{i}\nu\beta/2}\right)^{M}\nonumber \\
 & \times\int\exp\left\{ -\int\left[\mathbf{z}^{\dagger}\mathrm{d}\mathbf{z}+h\left(\mathbf{z}^{\dagger},\mathbf{z}\right)\mathrm{d}\tau\right]\right\} \mathrm{d}\mu_{W}^{\nu}\left(\mathbf{z}^{\dagger},\mathbf{z}\right).\label{Klaud}
\end{align}

Similarly to Eq. (\ref{main result-2}), Eq. (\ref{Klaud}) involves
truly continuous trajectories due to the presence of Wiener measure.
However, each trajectory is non-differentiable. Construction given
by Eq. (\ref{main result-2}) is based on smooth trajectories of class
$C^{\infty}$. Its drawback is that its real-time analog would be
only conditionally convergent, while Eq. (\ref{real-time-Klaud})
is on purpose absolutely convergent.

Since Eqs. (\ref{main result-2}) and (\ref{Klaud}) use different
Hamiltonian representations, it is a subjective opinion (to some extent)
that Weyl ordering is superior. Actually, many constructions are possible
and one needs to identify the most useful for a given purpose. The
one provided in this paper is certainly suitable for manipulations
performed in the Matsubara frequency domain.

\section{Implications and conclusion}\label{sec:Implications-and-conclusion}

The derived result can be widely used for treating bosonic many body
systems. It provides a clear instruction on how to use continuous
paths without contradictions and without distorting the partition
function. Additionally, the entire process of derivation throws some
light on the general nature of subtleties arising around taking continuum
limits and the role of renormalization in it. The key findings can
be captured by two rules: Hamiltonian should be Weyl-ordered and a
UV (high-frequency $\omega_{b}$) cut-off should be applied, which
finally should tend to infinity. Existing results, which treated order
and cut-offs carelessly, can be improved, while future works employing
path integrals can be done with more confidence.

Usually, Weyl-ordering typical Hamiltonian (e. g. Bose-Hubbard model)
does not change their structure, but modifies the parameters involved
by creating their linear combinations. While this does not influence
qualitative physics of the model, has a huge impact on the obtained
quantitative results.

\appendix

\section{Calculation of the normalizing factor}\label{sec:Calculation-of-the}

Applying the first recursive formula from Eq. (\ref{group flow})
and noting that $c_{B,B}=2^{\left(N-1\right)M}$ gives
\begin{equation}
c_{B,b}=2^{\left(N-1\right)M}\prod_{B^{\prime}=b+1}^{B}\left(4\tan^{2}\frac{\omega_{B^{\prime}}}{2}\right)^{-M}.\label{cBb}
\end{equation}
The emerged product can be handled in the limit $1\ll b\ll B$ using
similar tricks as that used to establish Eq. (\ref{prod identity}).
Additionally, it is helpful to assume $b^{3}\ll B^{2}$.

First, we note that
\begin{equation}
\frac{z^{N}-1}{z-1}=\prod_{\omega\neq0}\left(z-e^{\mathrm{i}\omega}\right).
\end{equation}
Then, taking a limit $z\rightarrow1$ produces
\begin{equation}
N=\prod_{\omega\neq0}\left(1-e^{\mathrm{i}\omega}\right).
\end{equation}
Now, we focus on
\begin{align}
\prod_{\left|n\right|>b}\frac{1}{2}\left(1+e^{\mathrm{i}\omega_{n}}\right) & =\frac{\prod_{n}\frac{1}{2}\left(1+e^{\mathrm{i}\omega_{n}}\right)}{\prod_{\left|n\right|\leq b}\frac{1}{2}\left(1+e^{\mathrm{i}\omega_{n}}\right)}\nonumber \\
 & =2^{2b}\frac{\prod_{n}\frac{1}{2}\left(1+e^{\mathrm{i}\omega_{n}}\right)}{\prod_{1\leq n\leq b}\left(2+2\cos\omega_{n}\right)}\nonumber \\
 & \cong\frac{\prod_{n}\frac{1}{2}\left(1+e^{\mathrm{i}\omega_{n}}\right)}{\prod_{1\leq n\leq b}\left(1-\frac{1}{4}\omega_{n}^{2}\right)}\nonumber \\
 & \cong\prod_{n}\frac{1}{2}\left(1+e^{\mathrm{i}\omega_{n}}\right)=2^{1-N}.
\end{align}
The weird assumption $b^{3}\ll B^{2}$ was used to state that $\prod_{1\leq n\leq b}\left(1-\frac{1}{4}\omega_{n}^{2}\right)\rightarrow1$.
Similarly:
\begin{align}
\prod_{\left|n\right|>b}\frac{1}{2}\left(1-e^{\mathrm{i}\omega_{n}}\right) & =\frac{\prod_{n\neq0}\frac{1}{2}\left(1-e^{\mathrm{i}\omega_{n}}\right)}{\prod_{0\neq\left|n\right|\leq b}\frac{1}{2}\left(1-e^{\mathrm{i}\omega_{n}}\right)}\nonumber \\
 & =\frac{2^{b+1-N}N}{\prod_{1\leq n\leq b}\left(1-\cos\omega_{n}\right)}\nonumber \\
 & \cong\frac{2^{2b+1-N}N}{\prod_{1\leq n\leq b}\omega_{n}^{2}}\nonumber \\
 & =\frac{2^{1-N}N^{2b+1}}{\pi^{2b}\left(\prod_{1\leq n\leq b}n\right)^{2}}\nonumber \\
 & =\frac{2^{1-N}N^{2b+1}}{\pi^{2b}\left(b!\right)^{2}}.
\end{align}
The prefactor in front of Eq. (\ref{main result}) can be finally
evaluated by joining (\ref{cBb}) with the derived product identities:
\begin{align}
 & \mathrm{prefactor}^{1/M}=\nonumber \\
 & =\left(\frac{N}{\beta}\right)^{2b+1}c_{B,b}^{1/M}\nonumber \\
 & =\left(\frac{N}{\beta}\right)^{2b+1}2^{\left(N-1\right)}\prod_{n>b}\left(4\tan^{2}\frac{\omega_{n}}{2}\right)^{-1}\nonumber \\
 & =\left(\frac{N}{\beta}\right)^{2b+1}2^{\left(N-1\right)}\prod_{\left|n\right|>b}\mathrm{i}\left(2\tan\frac{\omega_{n}}{2}\right)^{-1}\nonumber \\
 & =\left(\frac{N}{\beta}\right)^{2b+1}2^{\left(N-1\right)}2^{-\left[N-\left(2b+1\right)\right]}\frac{\prod_{\left|n\right|>b}\frac{1}{2}\left(1+e^{\mathrm{i}\omega_{n}}\right)}{\prod_{\left|n\right|>b}\frac{1}{2}\left(1-e^{\mathrm{i}\omega_{n}}\right)}\nonumber \\
 & =\beta^{-\left(2b+1\right)}\left(2\pi\right)^{2b}\left(b!\right)^{2}.
\end{align}

\bibliographystyle{unsrt}
\bibliography{bib}

\end{document}